\documentclass[fleqn,usenatbib]{mnras}

\usepackage[utf8]{inputenc}
\usepackage{enumitem}
\usepackage{graphicx}
\usepackage{hyperref}
\usepackage{mathtools}
\usepackage{comment}
\usepackage{amsmath}
\usepackage{ amssymb }
\usepackage{lipsum} 
\usepackage{float}
\usepackage{cleveref}
\usepackage[caption=false]{subfig}
\usepackage{placeins}
\usepackage{xcolor}

\newcommand{\lbgy}{C_{\ell}^{gy}}
\newcommand{\dd}{{\rm d}}

\definecolor{notecolor}{rgb}{0.8,0,0}
\definecolor{OrangeRed}{HTML}{FF4500}
\definecolor{MidnightBlue}{HTML}{191970}
\definecolor{MyGreen}{rgb}{0.3,0.8,0.3}
\definecolor{YellowGreen}{HTML}{9ACD32}
\definecolor{Indigo}{HTML}{4B0082}

\makeatletter
\def\apptablenumbers{\global
\setcounter{table}{0}
\setcounter{figure}{0}
\setcounter{equation}{0}
\def\thetable{\thesection\the\c@table}%
\def\fnum@table{{\bf\tablename~\thetable}}%
\def\thefigure{\thesection\the\c@figure}%
}%

\title[Probing reionization with galaxy-$y$ correlations]{The correlation of high-redshift galaxies with the thermal Sunyaev-Zel'dovich effect traces reionization}

\author[E. J. Baxter, L. Weinberger et al.]{
Eric J.~Baxter,$^{1,2,3,4}$
Lewis Weinberger,$^{2,3}$
Martin Haehnelt,$^{2,3}$
Vid Ir\v{s}i\v{c},$^{5,3}$
\newauthor
Girish Kulkarni,$^{6}$
Shivam Pandey,$^{7}$
and Anirban Roy$^{8}$
\\$^{1}$Institute for Astronomy, University of Hawai'i, 2680 Woodlawn Drive, Honolulu, HI 96822, USA
\\$^{2}$Institute of Astronomy, University of Cambridge, Madingley Road, Cambridge CB3 0HA, UK
\\$^{3}$Kavli Institute for Cosmology, University of Cambridge, Cambridge CB3 0HA, UK
\\$^{4}$Department of Applied Mathematics and Theoretical Physics, University of Cambridge, Cambridge CB3 0WA, UK
\\$^{5}$Cavendish Laboratory, University of Cambridge, 19 J. J. Thomson Ave., Cambridge CB3 0HE, UK
\\$^{6}$Department of Theoretical Physics, Tata Institute of Fundamental Research, Homi Bhabha Road, Mumbai 400005, India
\\$^{7}$Department of Physics and Astronomy, University of Pennsylvania, Philadelphia, PA 19104, USA
\\$^{8}$Department of Astronomy, Cornell University, Ithaca, NY 14853, USA
}

\begin{document}
\label{firstpage}

\pagerange{\pageref{firstpage}--\pageref{lastpage}}
\maketitle

\begin{abstract}
We explore a potential new probe of reionization: the cross-correlation of high-redshift galaxies with maps of the thermal Sunyaev-Zel'dovich (tSZ) effect.  We consider two types of high redshift galaxies: Lyman break galaxies (LBGs) and Lyman-$\alpha$ emitters (LAEs).  LBGs and LAEs will be detected in large numbers at high redshift ($z \approx$ 4 -- 7) by ongoing and future surveys.  We consider a future LBG sample from The Rubin Observatory Legacy Survey of Space and Time (LSST), and a selection of LAEs modelled after the Subaru SILVERRUSH program, but covering a much larger sky fraction.  The tSZ effect is sensitive to a line-of-sight integral of the ionized gas pressure, and is measured across large patches of sky using multi-frequency CMB surveys.  We consider forecast tSZ maps from CMB Stage 4 and more futuristic observations.  Using a suite of hydrodynamical simulations, we show that LBGs and LAEs are correlated with the tSZ signal from reionization.  The cross-spectra between LBGs/LAEs with tSZ maps contain information about the reionization history of the Universe, such as the distribution of bubble sizes, and could be used to directly measure the timing of reionization. The amplitude of the signal is small, however, and its detectability is hindered by low-redshift contributions to tSZ maps and by instrumental noise.  If the low-redshift contribution to the observed tSZ signal is suppressed by masking of massive halos, a combination of overlapping futuristic CMB and galaxy surveys could probe this signal.
\end{abstract} 


\section{Introduction}

Reionization impacts the observed cosmic microwave background (CMB) in several ways.  For one, CMB photons may Thompson scatter with the free electrons released by reionization.   This process impacts the observed temperature and polarization power spectra, enabling measurement of the optical depth to the last scattering surface, $\tau$.  A constraint on $\tau$ in turn constrains the timing of reionization, since an earlier onset will increase the volume of ionized gas that CMB photons must traverse, thereby increasing $\tau$ (for a review see \citealt{Reichardt:2016}).  CMB observations also constrain
reionization via the kinematic Sunyaev-Zel'dovich (kSZ) effect \citep{kSZ}.  The kSZ is a Doppler shift imparted to CMB photons by inverse Compton scattering with ionized gas that has bulk velocity relative to the CMB rest frame (for a review see \citealt{Birkinshaw:1999}). The inhomogeneously ionized Universe during reionization leads to a ``patchy" kSZ signal that can be observed by CMB experiments \citep{Gruzinov:1998}. Measurements of this patchy signal constrain the duration of reionization \citep[e.g.][]{Reichardt:2020}.

Reionization should also leave an impact in the CMB via the {\it thermal} Sunyaev-Zel'dovich (tSZ) effect \citep{SZ}, which results from inverse Compton scattering of CMB photons with hot ionized gas.  This process leads to a spectral distortion in the CMB that is sensitive to Compton-$y$, which is in turn proportional to a line-of-sight integral of the ionized gas pressure:
\begin{equation}
y = \frac{\sigma_T}{m_e c^2}\int_0^{\infty} \dd l \, P_e(l),
\label{eq:ydef}
\end{equation}
where $P_e(l)$ is the electron pressure and $l$ is the line of sight distance.  The same bubbles of warm ionized gas that lead to the patchy kSZ signal should also lead to an inhomogenous $y$ signal.  Since $y$ scales like $k_B T/m_e c^2$, where $T$ is the gas temperature,  while the kSZ signal scales like $v/c$, where $v$ is the gas bulk velocity, the relative amplitude of the tSZ signal to the kSZ signal is roughly $Tk_B/m_e c v$.   The temperature inside the reionized bubbles of gas is roughly $T\sim 10^4\,{\rm K}$, and a characteristic velocity is $v \sim 200\,{\rm km}/{\rm s}$.  Substituting these values,  the thermal SZ signal from reionization is expected to be about two orders of magnitude smaller than the kinematic signal.  However, the $y$ signal carries additional information about the thermodynamics of the gas that could in principle be used to constrain reionization models.  Furthermore, unlike the kSZ, the tSZ leads to a distinct spectral distortion in the CMB, making it more straightforward to estimate from multifrequency CMB observations than the blackbody-preserving kSZ signal.

The $y$ signal is sensitive to all gas between the observer and the last scattering surface.  Because the hot gas in halos at low redshift makes a dominant contribution to the integrated pressure, measurements of the $y$ power spectrum are essentially only sensitive to the low-redshift gas that lives in the most massive dark matter halos \citep[e.g.][]{Makiya:2018}.  Extracting  the high-redshift contribution to the $y$ power spectrum would require modeling (and subtracting) the low-redshift contribution with extremely high accuracy.

However, by cross-correlating maps of Compton-$y$ with a tracer of the large scale structure at high redshift, such as galaxies, one can effectively isolate the contributions to $y$ from the high-redshift Universe.  In this case, the low-redshift contribution acts solely as a noise source.  A detection of this correlation as a function of redshift has the potential to directly probe how the Universe reionized over cosmic time.  

One way to select large populations of high-redshift galaxies for this purpose is via wide-field photometric surveys.  Ongoing (e.g. the GOLDRUSH catalog from HSC, \citealt{ONO18}) and future (e.g. the Vera C. Rubin Observatory Legacy Survey of Space and Time, LSST\footnote{\url{http://www.lsst.org}}, \citealt{LSST}; WFIRST, \citealt{WFIRST}; EUCLID, \citealt{EUCLID}) deep imaging surveys can employ so-called dropout methods to select high redshift galaxies.  As the Lyman break feature in a galaxy spectrum passes through the optical filter bands, the galaxy will disappear (i.e. dropout) in the bluer bands.  Simple color and flux cuts can be used to select populations of these Lyman break galaxies (LBGs) localized in fairly narrow redshift intervals.  For instance, $g$-band dropouts are localized at around $z \sim 4$, $r$-band dropouts are localized around $z\sim 5$, etc.  Large samples of dropout-selected galaxies are expected from LSST and other future imaging surveys \citep{Wilson:2019}.  

Another class of galaxies that can be relatively easily identified at high redshift are Lyman-$\alpha$ emitters (LAEs). Narrow band searches can be used to identify the redshifted Lyman-$\alpha$ emission from these galaxies out to $z \sim 7$.  While the largest samples of LAEs to date have been derived from fairly small area surveys  --- such as the $\sim 20\,{\rm deg}^2$ SILVERRUSH survey using the Subaru telescope \citep{OUCHI18} --- future samples may be obtained from much larger sky areas using e.g. SPHEREx \citep{SPHEREX}. 

\begin{figure*}
    \centering
    \includegraphics[width=\textwidth]{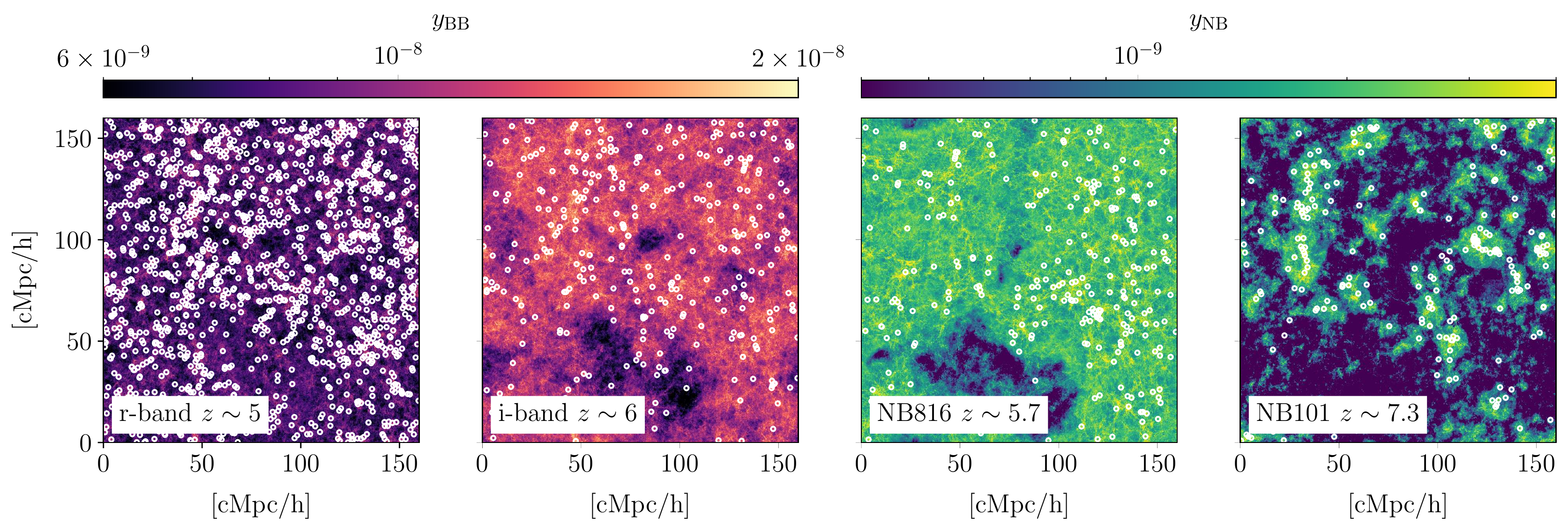}
    \caption{Simulated galaxy populations and the contribution to the
      Compton-$y$ signal from within the corresponding galaxy survey
      band. From left to right we show the slices at $z=5.000, 5.946,
       5.756$ and $7.444$, with mock LBG and LAE positions indicated by
      white circles. The left two panels show the simulated $y$-signal from 
      integrating across the mock broadbands, $y_{\rm BB}$, corresponding to r- and i-band dropout galaxy selections; 
      the mock LBG population are overlayed as white circles. The right panels 
      similarly show the simulated $y$-signal from integrating across the mock NB816 and 
      NB101 narrowbands, $y_{\rm NB}$, with the LAE population overlayed as white 
      circles. Both galaxy populations are highly spatially correlated with the
      $y$-signal.}
    \label{fig:y_maps}
\end{figure*}

In this work, we explore the cross-correlation signal between high redshift galaxies (both LBGs and LAEs) and Compton-$y$ using the Sherwood simulation suite \citep{BOLTON17}, a set of IGM-focused hydrodynamical simulations.  We show that the galaxy-$y$ correlation provides a distinctive signal of reionization.  We then determine whether future surveys have a hope of detecting this signal.   

Several recent works \citep[e.g.][]{Sobacchi:2016, Kubota:2018} have considered the related possibility of probing reionization with measurement of the correlation between high-redshift galaxies, such as LAEs, and maps of the redshifted 21-cm signal from neutral hydrogen.  The correlation considered here --- between similar galaxies and $y$ --- is complementary to the 21-cm correlation, since it provides additional information about the gas pressure in the ionized bubbles.  Additionally, the estimation of the $y$ and 21-cm signals from data must contend with different observational challenges.  Brightness temperature fluctuations due to galactic synchrotron and extragalactic point sources are orders of magnitude larger than the expected 21-cm signal \citep[e.g.][]{Shaver:1999}.  In the case of $y$, the main observational challenges are instrument noise and contributions to $y$ from low-redshift structure, as we discuss in more detail below.

The paper is organized as follows.  In \S\ref{sec:simulations} we describe the simulations used to compute the cross-correlation between galaxies and Compton-$y$; in \S\ref{sec:methods} we described the methodology for forecasting constraints on these cross-correlations from future surveys; our results are presented in \S\ref{sec:results}, and we conclude in \S\ref{sec:discussion}.  In Appendix~\ref{sec:feedback} we explore how our results are impacted by the feedback modelling within our simulations, whilst in Appendix~\ref{sec:calculation} we provide simple analytical estimates for the amplitude of the $y$ signal from reionized bubbles.

\section{Simulations}
\label{sec:simulations}

We employ the same simulation setup as in \citet{ROY20}, using the Sherwood simulation suite \citep{BOLTON17} to model the LBG, LAE and $y$ signals. This suite of cosmological hydrodynamical simulations was designed to capture the evolution of the low-density inter-galactic medium (IGM). 
We employ the delayed reionization history that has recently been found to explain Ly$\alpha$ forest opacity fluctuations \citep{KULKARNI18, KEATING19, NASIR19}, the high-redshift LAE luminosity function evolution \citep{WEINBERGER19}, and the low Thomson scattering optical depth to the CMB \citep{PLANCK18}.

\subsection{Reionizing the Sherwood simulations}
The Sherwood suite was run using the \textsc{P-Gadget-3} smoothed-particle hydrodynamics (SPH) code \citep{SPRINGEL05, SPRINGEL01}, with a range of simulation box sizes and resolutions. In this work we make use of the largest volume which has a side length of $L=160$ cMpc/h\footnote{We use the prefix c to denote comoving units, whilst p indicates proper units.} and particle number $N=2\times2048^3$, giving a dark matter particle mass of $M_{\rm DM} = 3.44\times 10^7 M_{\odot}/h$. The cosmological parameters were derived from the \citet{PLANCK13} results, including $h=0.678$, $\Omega_m = 0.308$, $\Omega_\Lambda = 0.692$, $\Omega_b = 0.0482$, $\sigma_8 = 0.829$, $n = 0.961$, and $Y_{\mathrm{He}} = 0.24$. To improve computational efficiency, the \texttt{QUICK\_LYA} star formation prescription \citep{VIEL04} was employed, in which cold dense gas ($T < 10^5$ K, $\Delta > 10^3$) is converted into star particles (see Appendix~\ref{sec:feedback} for discussion of the impact of this prescription on our results). From this simulation we use the SPH kernel to interpolate the hydrodynamic quantities onto a uniform grid at our target redshifts of interest $z=5.000,\: 5.756,\: 5.946,\: 6.604,\: 6.860,\: 7.444$. These redshifts overlap with the broadband $i$ and $r$ filters  (which can be used to identify high-redshift LBGs via dropout selection), and the narrowband NB816, NB921, NB973, and NB101 filters (which can be used to identify high-redshift LAEs).

We employ the late reionization history of \citet{KULKARNI18} by post-processing the simulations with the radiative transfer code \textsc{Aton} (for further details of the code, see \citealt{AUBERT10,AUBERT08}; for details of the simulation setup, see \citealt{KULKARNI18}).  We note that the shape and evolution of the cross-correlation signal is dependent on the reionization history (which controls the ionized bubble distribution, see Appendix~\ref{sec:calculation}) and hence by fixing this choice we are forecasting only for this specific case. As discussed further in \citet{ROY20} and \citet{WEINBERGER20}, our chosen reionization history is consistent with existing observations of the Lyman-$\alpha$ forest \citep{KULKARNI18, KEATING19, NASIR19}, LAE populations \citep{WEINBERGER19} and the CMB optical depth \citep{PLANCK18}.  We refer the interested reader to Figure 1 of \citet{KULKARNI18} for further details of this reionization history.

To push our forecasts to scales larger than our fiducial 160 cMpc/h simulation volume, we also employ a semi-numerical simulation with a side length of 1 cGpc \citep{ROY20}, run using 21cmFast \citep{MESINGER11}. This simulation lacks the spatial resolution to accurately model the smaller scale correlations between $y$ and the galaxy populations, however it is effective for predicting the large scale modes. We post-process this simulation with the same reionization history as the Sherwood simulation, using the calibrated excursion set method of \citet{CHOUDHURY15}. For further implementation details and a comparison of these simulations, we refer the reader to \citet{ROY20}.

While the bulk of our analysis focuses on results from numerical simulations, in Appendix~\ref{sec:calculation} we provide analytical estimates for the amplitude of the signal that are in good agreement with the simulation results.

\subsection{Modelling the $y$ signal}

It is useful to consider the contribution to the total $y$ from structure between redshifts $z_{\rm min}$ and $z_{\rm max}$.   Reframing Eq.~(\ref{eq:ydef}) as an integral over redshift (here using the \emph{comoving} electron pressure), we have
\begin{eqnarray}
y(z_{\rm min}, z_{\rm max}) &=& \frac{\sigma_T}{m_e c^2} \int_{z_{\rm min}}^{z_{\rm max}} \dd z \: (1+z)^2 \:\frac{\dd \chi}{\dd z}\: P_e(z) \nonumber \\ 
&\equiv& \int_{z_{\rm min}}^{z_{\rm max}} \dd z \:W_y(z) \: P_e(z)
\label{eq:y_integral}
\end{eqnarray}
where $\chi$ is comoving distance, and the window function is given by,
\begin{eqnarray}
W_y(z) \equiv  \frac{\sigma_T}{m_e c^2} \:(1+z)^2 \:\frac{\dd\chi}{\dd z}.
\end{eqnarray}
The $y$-signal measured by CMB experiments is sensitive to the full integral of the ionized gas pressure from to $z_{\rm min}=0$ to $z_{\rm max} = z_{\rm CMB} \sim 1100$ (see Eq.~\ref{eq:ydef}).  However the cross-correlation with a galaxy survey will pick out the contribution to $y$ from within the survey volume. In practice, we perform the integral in Eq.~\ref{eq:y_integral} across the redshift limits of the galaxy bands, and use the notation $y_{\rm BB}$ or $y_{\rm NB}$ (broadband and narrowband, respectively) for this ``band-contribution'' to the observed (total) $y$.  This procedure will capture the contributions to $y$ relevant for estimation of the cross-correlation signals.  Of course, the redshift distributions of the selected galaxies will not be exactly described by top-hat functions, so Eq.~\ref{eq:y_integral} is an approximation.  Furthermore, in actual data, there is the possibility of low-redshift interlopers contaminating the galaxy samples \citep[e.g.][]{Dunlop:2013}.  Given our focus on idealized forecasts, we postpone a careful consideration of these issues to future work.  Finally, Eq.~\ref{eq:y_integral} ignores contributions to $y$ that may be correlated with the selected galaxies, but that extend outside of the bands considered.  Given that in all cases, the bands are several times wider than the typical correlation length of the galaxy and $y$ fields, we expect these contributions to be small.  Our choice to ignore these contributions is motivated by the fact that uncorrelated $y$ in the simulated $y$ maps will add noise to the correlation measurements.

After interpolating the SPH particles and running the post-processing radiative transfer, we have grids of $x_{\rm e}$, $n_{\rm H}$ and $T$ which can be used to derive $P_e$. The simulated contribution to the $y$-signal from broadband and narrowband survey volumes is shown in Figure~\ref{fig:y_maps}. These maps were created by integrating Eq.~\ref{eq:y_integral} across the redshift range of the band. We note that we do not account for the lightcone effect here, which may introduce some inaccuracy for the broadband filters which span $\Delta z \gtrsim 1$.

\subsection{Modelling the LBG and LAE populations}
We use the galaxy modelling described in \citet{WEINBERGER19}. We will briefly summarise the details of this modelling here, and refer the interested reader to that work for further information. 

We start with the dark matter halo population of the simulation, which is found on-the-fly with a Friends-of-Friends finder. At $z=5$ the minimum and maximum halo masses are $M_h^{\rm min} = 2.3\times10^8$ M$_\odot/h$ and $M_h^{\rm max}=4.6\times10^{12}$ M$_\odot/h$, respectively, whilst at $z=7.444$ the maximum mass is $M_h^{\rm max}=2.7\times10^{12}$ M$_\odot/h$.  This halo population is abundance matched to the UV luminosity function from \citet{BOUWENS15} at $z\sim6$ with the duty cycle formalism from \citet{TRENTI10} in order to create a mock LBG population.  The resulting mapping from halo mass to broadband UV luminosity, $L_{\rm UV}(M_h)$, can be applied to the halo population at other redshifts to recover the observed evolution in the luminosity function. 

To generate the LAE population we find the subset of this LBG population with observable Ly$\alpha$ luminosities and equivalent widths. The Ly$\alpha$ properties of the LBG population are modelled statistically using the equivalent width distribution model of \citet{DIJKSTRA12}. We calculate the IGM transmission along single sightlines through the simulated IGM to each LAE in order to account for the effect of reionization\footnote{We note that we only model the LAE population in the 160 cMpc/h Sherwood simulation and not the 1 cGpc simulation, as the Ly$\alpha$ transmission calculation is sensitive to how well the simulation can resolve the small-scale self-shielded neutral hydrogen around halos.} \citep[for further details see][]{WEINBERGER18}.

When calculating correlation statistics we use the dimensionless overdensity,
\begin{eqnarray}
\delta_g^{2D} &=& \int_{z_{\rm min}}^{z_{\rm max}} \dd z \: \frac{n(z)}{\bar{n}} \: \delta^{3D}_g(z)  \\
&\equiv& \int_{z_{\rm min}}^{z_{\rm max}} \dd z \: W_g(z) \: \delta^{3D}_g(z),
\end{eqnarray}
where $n(z)$ is the redshift distribution of galaxies, and the galaxy window function is given by, 
\begin{eqnarray}
W_g(z) = \frac{n(z)}{\bar{n}}
\end{eqnarray}
Here we match the window function to the desired redshift distribution of both the broadband \citep{ONO18} and narrowband \citep{SHIBUYA18} filters. In order to forecast realistic galaxy shot noise we calibrate the luminosity selection such that we recover expected survey number densities, $\bar{n}$. For the mock LAE populations we match to the observed densities from the SILVERRUSH survey \citep{OUCHI18}, whilst for the LBG populations we match to forecast number densities from \citet{Wilson:2019}

\subsection{Correlation statistics}
\label{sec:correlation_stats}

Given the full 3D information content of the simulation, we first calculate the spatial power spectrum in 3D,
\begin{equation}
\langle \tilde{\delta}_{\rm g} (\mathbf{k_1})\tilde{P}_e (\mathbf{k_2}) \rangle = (2 \pi)^3 \delta_{D}(\mathbf{k_1} + \mathbf{k_2}) P_{\rm g P_e} (k),
\end{equation}%
where a tilde indicates the Fourier transformed field, and $P_{\rm g P_e} (k)$ is the cross-power spectrum between LAEs and the electron pressure. Using the Limber approximation \citep{LIMBER53, PEEBLES80} this can be projected to the 2D cross-power spectrum,
\begin{multline}
C_{\ell}^{g y} = \frac{\sigma_T}{m_e c^2}  \int \dd z \: (1+z)^2 \: \frac{n(z)}{\bar{n}} \: \frac{1}{\chi^2(z)}\\
\times P_{g P_e}\left( k = \frac{\ell + 1/2}{\chi(z)};z \right),
\end{multline}
where we have further assumed that the comoving angular diameter distance $d_A = \chi$, valid in the limit of a flat universe. We construct the auto-power spectra similarly.

\section{Forecasting methodology}
\label{sec:methods}

We now forecast the ability of CMB and galaxy surveys to measure $\lbgy$.  Measurement of $\lbgy$ must contend with three sources of noise: (1) shot noise from the galaxy survey, (2) instrument and foreground noise in the estimated Compton-$y$ maps, and (3) the $y$ contributions from structure at low redshift (i.e. after reionization) that is uncorrelated with the signal at high redshift.  For future surveys, the $y$ noise contribution from low redshift structure can be comparable to  or greater than that of instrument and foreground noise.  However, as we discuss in more detail below, the contribution from low-redshift structure can be suppressed by masking detected halos at low redshift.

The covariance of the $\lbgy$ cross-spectrum in principle includes four-point combinations of the galaxy and Compton-$y$ fields.  However, because the Compton-$y$ map is dominated by low-$z$ $y$ and noise --- which are both uncorrelated with the reionization signal --- we can ignore the connected part of the four-point function and consider only the Gaussian limit.  In this limit, the error on the measured $\lbgy$ is given by 
\begin{multline}
\sigma^2(C_{\ell_i}^{gy}) = \frac{1}{(2\ell + 1) \Delta \ell f_{\rm sky}}\\
\left[ \left( C_{\ell}^{yy-{\rm lowz}} + N_{\ell}^{yy} \right)\left(C_{\ell}^{gg} + \frac{1}{\bar{n}} \right) +  \left(\lbgy \right)^2 \right],
\end{multline}
where $\ell_i$ labels the $\ell$ bin, $\Delta \ell$ is the width of the bins, $f_{\rm sky}$ is the observed sky fraction, $C_{\ell}^{yy-{\rm lowz}}$ is the low-redshift contribution to the $y$ auto-spectrum, $N_{\ell}^{yy}$  is the noise and foreground contribution to the $y$  auto-spectrum,  and $C_{\ell}^{gg}$ is the galaxy auto-spectrum.  We ignore the contribution to the total $y$ auto-power from high redshift $y$, since this contribution is subdominant compared to the instrumental noise and low-redshift contributions. \citet{Peng_Oh_2003} found the $y$ auto-spectrum sourced by reionization to be roughly an order of magnitude below the low redshift contribution using a model for which the mean $y$ contributed by reionization is $\langle y_{\rm reion} \rangle \sim 3.6\times 10^{-6}$.  A more recent calculation by \citet{Hill_2015} suggests $\langle y_{\rm reion} \rangle \sim 10^{-7}$, implying that the $y$ auto-spectrum from reionization is three to four orders of magnitude lower than that from low-redshift structure (see also Namikawa et al., in prep).

We compute the $C_{\ell}^{gg}$ and $C_{\ell}^{gy}$ terms directly from the simulations, as described in \S\ref{sec:correlation_stats}.  We discuss the computation of  $C_{\ell}^{yy-{\rm lowz}}$ and $N_{\ell}^{yy}$ in the following section.

\begin{figure}
    \centering
    \includegraphics[scale=0.4]{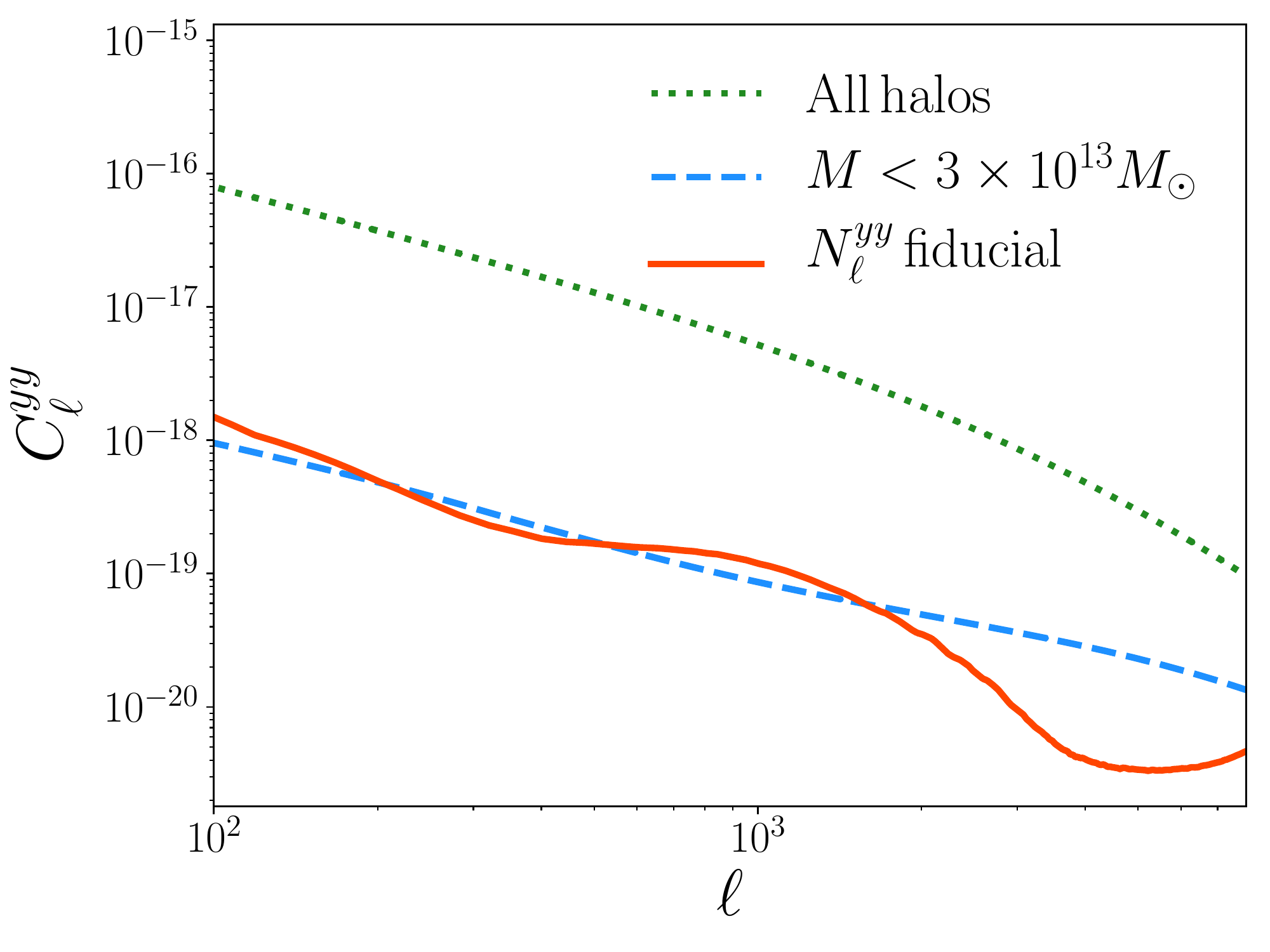}
    \caption{The Compton-$y$ auto-spectrum before (green dotted) and after (blue dashed) masking halos with $M  > 3\times  10^{13}\,M_{\odot}$.   Masking massive halos significantly reduces the $y$ auto-spectrum, which acts as a source of noise for the measurement of the cross-correlation between high-redshift galaxies and $y$.  Also shown is the fiducial noise level assumed for forecasts (red solid curve), which is a factor of ten lower than that expected for CMB-S4.}
    \label{fig:y_noise}
\end{figure}

\begin{figure*}
    \centering
    \includegraphics[scale=0.5]{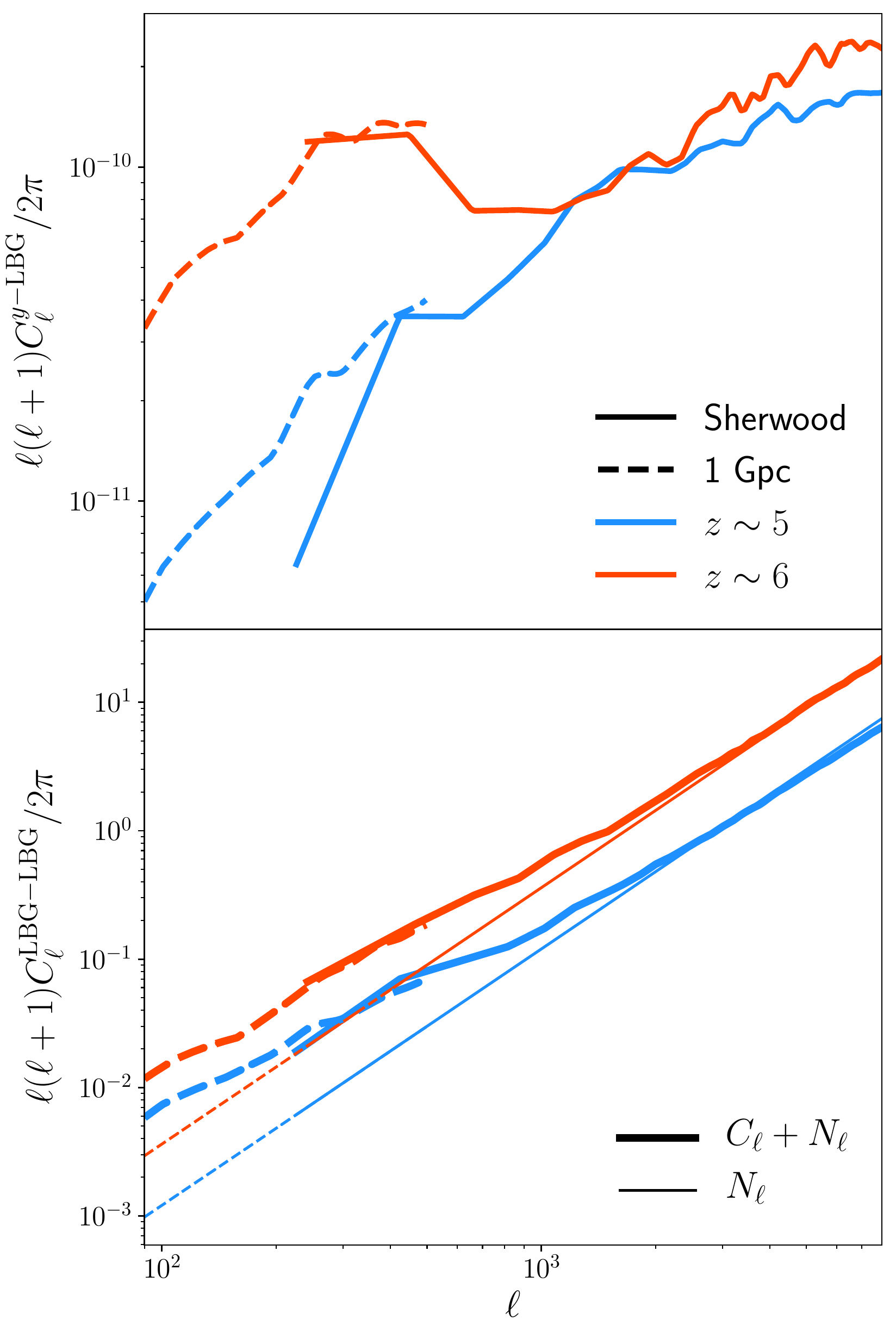}
    \includegraphics[scale=0.5]{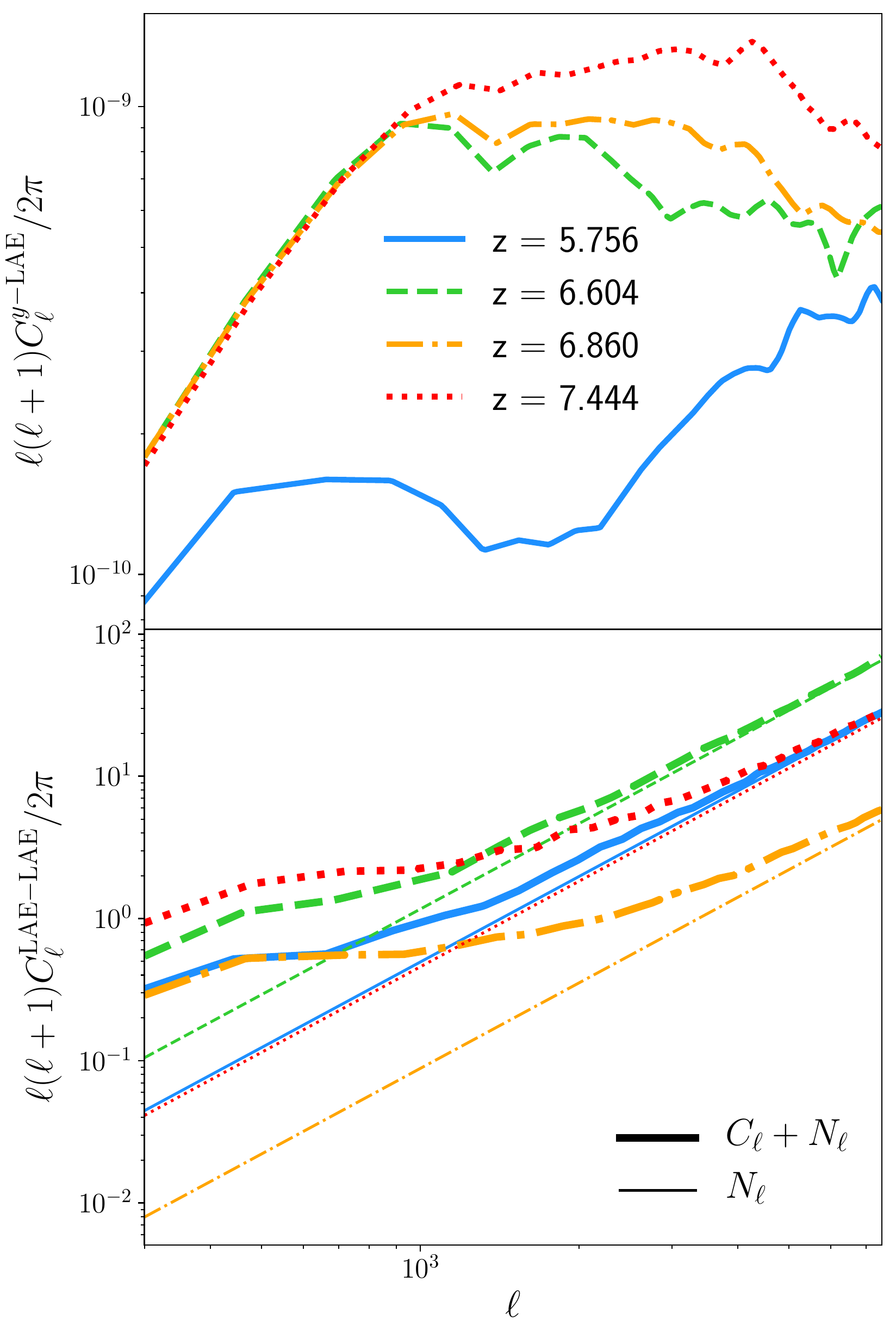}
    \caption{The galaxy-$y$ cross-spectra (top  panels) and galaxy clustering auto-spectra (bottom panels) computed  from the simulations. The left panels show the results for LBGs, while the right panels show the results for LAEs.  For the LBG results, we show power spectra computed for the 160 cMpc/h Sherwood simulation with solid lines, and power spectra for the 1 cGpc simulation with dashed lines. In the bottom panels, the thick curves correspond to the total (signal and noise) galaxy auto-power, while the thin lines correspond to the shot noise for each galaxy sample.
    }
    \label{fig:spectra}
\end{figure*}

\subsection{Noise in Compton-$y$ maps}
\label{sec:noise}

We compute the low redshift (i.e. post-reionization) contribution to the $y$ auto-power, $C_{\ell}^{yy-{\rm lowz}}$, using the halo model formalism.  Details of the halo model calculation for the $y$ auto-power can be found in e.g. \citet{Vikram:2017}, \citet{Hill:2018} and \citet{Pandey:2019}.  We exactly follow the methodology described in \citet{Pandey:2019}.  In this model, all gas in the Universe resides in halos.  The $y$ auto-spectrum can then be computed when the pressure profiles of halos of a given mass and redshift are known, and when the clustering of the halos is specified.   We assume the halo pressure profiles are specified by the fits to hydrodynamical simulations from \citet{Battaglia:2012}.  In these fits, the halo pressure profile is described by a generalized Navarro-Frenk-White profile \citep{NFW},
\begin{eqnarray}
P(x  =  r/R_{200c}) = P_0(x/x_c)^{\gamma}(1  + (x/x_c)^{\lambda})^{-\beta},
\end{eqnarray}
where $r$ is the distance to the halo center, $R_{200c}$ is the halo radius, and $x_0$, $\beta$, $\lambda$ and $\gamma$ are model parameters varied in the fits.  The clustering of halos is described by a linear bias model.  We show the computed $C_{\ell}^{yy-{\rm lowz}}$ curve as the green dotted line in Fig.~\ref{fig:y_noise}.  

The integrated Compton-$y$ signal from a halo of mass $M$ scales roughly as $M^{5/3}$; as a result, the most massive halos make a large contribution to the total $y$ power despite their low abundance.  Fig.~3 from \citet{Makiya:2018} nicely illustrates that the bulk of the contributions to the $y$ auto-spectrum come from halos with $M \sim 10^{15}\,M_{\odot}$.  Consequently, if massive  halos can be detected and masked, the total $C_{\ell}^{yy-{\rm lowz}}$ can be reduced significantly.   In Fig.~\ref{fig:y_noise} we show the impact on $C_{\ell}^{yy-{\rm lowz}}$ when halos with $M > M_{\rm threshold} = 3\times 10^{13}\,M_{\odot}$ are removed.  This mass scale corresponds roughly to large galaxy groups, which will be detected over large patches of the sky by e.g. LSST \citep{LSST}.  The resultant reduction in power is nearly two orders of magnitude for $\ell \lesssim 5000$.  

Although masking low-redshift halos will not bias the high-redshift correlation measurements (since the low-redshift halos are uncorrelated with those at high-redshift), it will reduce available sky area and thereby degrade the signal to noise.  However, even for the aggressive masking of halos with $M > M_{\rm threshold} = 3\times 10^{13}\,M_{\odot}$, the reduction in available sky area is only about 10\% if one assumes that halos are masked with a disk of radius 1 arcminute (for the most massive halos, this level of masking will not be sufficient, but the number of such halos is small).  One can significantly reduce the masked fraction at the cost of slightly enhanced low-redshift $y$ noise power by increasing $M_{\rm threshold}$ by a small amount.  A future analysis with actual data could determine an optimal $M_{\rm threshold}$ from a signal-to-noise standpoint.  For simplicity, we will ignore the reduction in $f_{\rm sky}$ caused by masking below.

As we will show below, the expected signal-to-noise for the galaxy-$y$ correlation measurements is low, even for futuristic surveys.  We therefore make forecasts for both CMB Stage 4 (CMB-S4; \citealt{CMBS4}), and for a more futuristic survey (which we refer to as CMB-S5) that has a factor of ten lower noise in Compton-$y$.  The $y$ noise forecasts for CMB-S4 are taken from \citet{CMBS4_DSR}.  Since the noise in the estimate of $y$ is very roughly proportional to the temperature noise, our estimated $y$ noise for CMB-S5 is about a factor of four better than that expected for Probe of Inflation and Cosmic Origins (PICO) \citep{PICO}.  While adopting such a low noise level is clearly optimistic, taking this route makes it clear how much improvement relative to CMB-S4 is needed in order to obtain interesting constraints on reionization from these measurements.  The fiducial noise spectrum (for CMB-S5) is shown as the red solid curve in Fig.~\ref{fig:y_noise}.  We ignore potential issues of contamination in the $y$ maps (such as from the cosmic infrared background), but return briefly to this issue in \S\ref{sec:discussion}.

We will adopt $f_{\rm sky} = 0.5$ for all forecasts.  This is a reasonable assumption if one considers CMB-S4 or a space-based CMB survey (like PICO) correlated with galaxies identified from a wide-field survey like LSST.  Dropout techniques should make selection of LBGs at high-redshift from such wide-area surveys feasible.  Using $f_{\rm sky} = 0.5$ is much more optimistic for the LAE-based forecasts, considering that current LAE surveys cover at most $\sim 20\,{\rm deg}^2$ \citep{ONO18}.  Expanded sky coverage is expected in the future from missions such as SPHEREx \citep{SPHEREX}, which will enable intensity mapping of LAEs over its roughly $200\,{\rm deg}^2$ deep fields.  At the very least, our use of $f_{\rm sky} = 0.5$  makes it easier to compare the LBG and LAE forecasts.  Metal-line intensity mapping, such as [{\sc C ii}] \citep{GONG12} measured by the CONCERTO survey \citep{LAGACHE18}, may also be a useful tracer of high-redshift large scale structure \citep{DUMITRU19}. We expect the $y$-[{\sc C ii}] cross-correlation to look similar to $y$-LBG, however we leave survey forecasts of that signal to future work.

\begin{figure}
    \centering
    \includegraphics[width=\columnwidth]{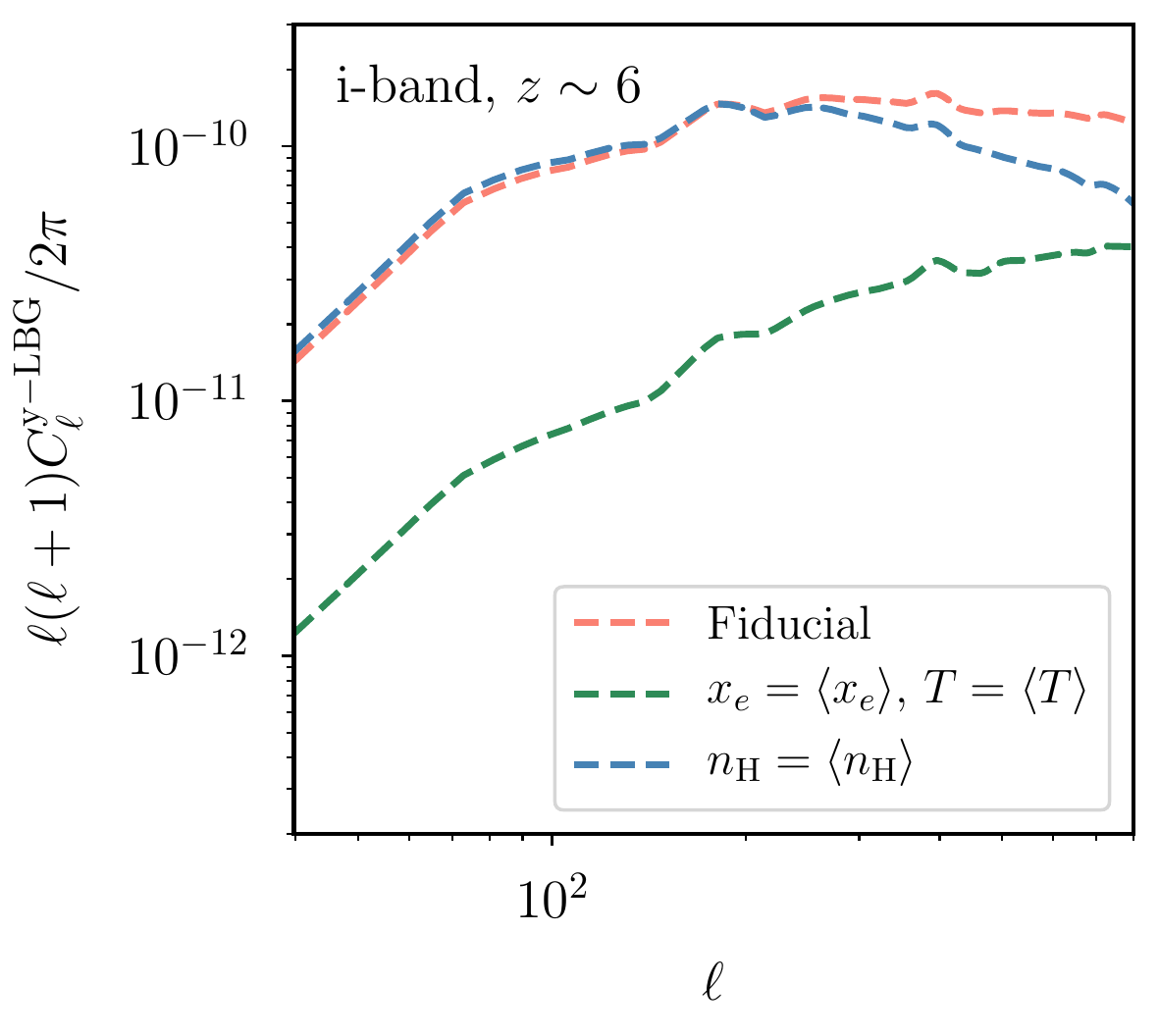}
    \caption{The impact on the large-scale LBG-$y$ cross-power of nulling  fluctuations in the density field (blue), and fluctuations in the ionization and temperature fields (green).  Results are shown for the 1 Gpc simulation with galaxies corresponding to the $z\sim 6$ i-band dropout selection. Setting the ionization and temperature fields to their mean values reduces the cross-power by roughly an order of magnitude, whereas setting the density to the mean value only deviates from the fiducial case (pink curve) on smaller scales $\ell >400$.  These results suggest that the LBG-$y$ cross-power on large scales is driven by fluctuations in the ionization and temperature fields.}
    \label{fig:source_of_power}
\end{figure}

\section{Results}
\label{sec:results}

\begin{figure*}
    \centering
    \includegraphics[scale=0.6]{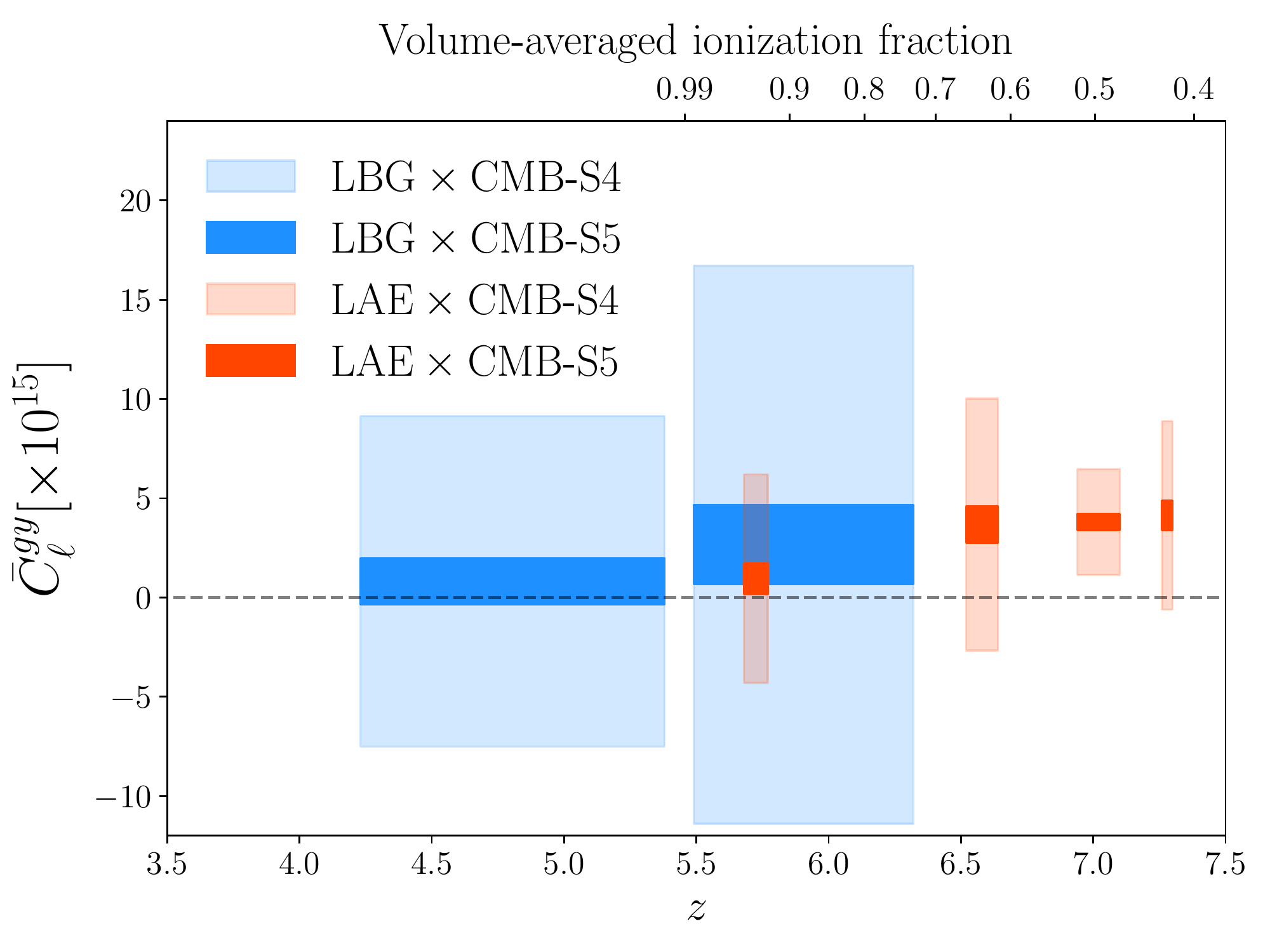}
    \caption{  Forecast constraints on the amplitude of the $\lbgy$ correlation as a function of $z$ for LAEs (red) and LBGs (blue).   The $y$-axis shows the average $\lbgy$ over the range $\ell = 300$ to $1000$ ($\ell = 300$ to $3000)$ for LBGs (LAEs), where most of the reionization signal is located.  The horizontal extent of each band indicates the redshift range of the corresponding galaxy selection, while the vertical extent represents the $1\sigma$ uncertainty on the cross-power amplitude.   The fainter shaded regions correspond to forecasts for $y$-maps from CMB-S4, while the darker regions correspond to a more futuristic survey with ten times lower $y$ noise than CMB-S4, i.e. what we call CMB-S5.   We have assumed that halos with $M > 3\times 10^{13} M_{\odot}$ have been identified and masked in the $y$-maps to reduce noise from low-redshift structure.  The top $x$-axis shows the volume-averaged ionization fraction at the redshifts indicated by the bottom $x$-axis.}
    \label{fig:gy_error}
\end{figure*}

The galaxy-$y$ and galaxy-galaxy spectra measured from the simulations corresponding to Fig.~\ref{fig:y_maps} are shown in the top panel of Fig.~\ref{fig:spectra}, with the left (right) panels corresponding to the results for LBGs (LAEs).  As apparent already at the map level in Fig.~\ref{fig:y_maps}, the LAEs and LBGs are correlated with Compton-$y$. This is expected, as the pressure of the gas and the ionization field --- which together yield the Compton-$y$ field --- are correlated with large scale structure, as are the LBGs/LAEs.

Redshift evolution in the galaxy-$y$ cross-spectra is also apparent in Fig.~\ref{fig:spectra}.  At high redshift ($z  \gtrsim 6$), the cross-spectra show an excess of power relative to the cross-spectra at lower redshift.   This excess is due to additional power in the Compton-$y$ field due to the presence of ionized bubbles at high redshift.  We provide a rough calculation of the expected excess cross-power in  Appendix~\ref{sec:calculation}.  As seen most clearly in the top right panel of Fig.~\ref{fig:spectra}, the bubbles grow larger with decreasing redshift, causing a shift in the peak of the $y$-galaxy cross-spectra to larger scales (lower $\ell$). Between $z \sim 7.4$ to $z \sim  6.6$ the position of the high-$\ell$ falloff in power moves from roughly $\ell \sim 5000$ to $\ell \sim 2000$. The cosmological scale factor changes by only a factor of $\sim 10\%$  over this period, indicating that the bubbles are indeed growing in physical size.

To determine whether the variations in the pressure or the ionization fields are driving the signal seen in Fig.~\ref{fig:spectra}, in Fig.~\ref{fig:source_of_power} we show the impact of replacing the ionization field and temperature field, or the density field, with constant values.  For simplicity, we plot only the cross-spectra with the LBG $i$-band sample.  As seen in the figure, setting the density field to its mean value (blue dashed curve) has little impact on the cross-power at $\ell \sim 200-300$, where the feature seen in the left top panel of Fig.~\ref{fig:spectra} peaks.  On the other hand, setting the ionization and temperature fields to their mean values has a large impact on the measured cross-power.  This suggests that the large scale power seen in the $\lbgy$ cross-spectrum is dominated by fluctuations in the ionization and temperature fields.   Similar results hold for the cross-spectra with the LAEs, albeit on smaller scales at high redshift.

The bottom panels of Fig.~\ref{fig:spectra} show the galaxy auto-power spectra (thick curves) including the shot noise contribution (thin curves).  For the most part, over the range of scales of interest for probing reionization, the galaxy clustering power dominates over the shot noise.  Consequently, increasing the galaxy density is not expected to result in significant increases in the signal-to-noise of the galaxy-$y$ correlations. We note that the galaxy auto-power spectra are dependent on the survey selection.  Changing the selection will change the galaxy bias, as well as the level of shot noise.  The choice of LAE luminosity and equivalent width cutoffs for the different narrowband surveys results in the different shot noise levels seen in the lower right panel of Fig.~\ref{fig:spectra} (see Table 1 of \citealt{WEINBERGER20} for further details of the LAE selection).

In Fig.~\ref{fig:gy_error} we show the forecast errorbars on the amplitude of the $y$-LBG cross-power spectra (for both LAEs and LBGs)  as a function of redshift.   We have defined this amplitude as the mean value of $\lbgy$  over the range of $\ell \in [300,1000]$ for LBGs and $\ell \in [300,3000]$ for LAEs, where most of the reionization signal is located.   For the fiducial LAE forecast (solid red), we project that this amplitude can be measured with signal-to-noise of roughy $1.2$, $4.0$, $9.2$ and $5.6\sigma$ in order of increasing redshift bin; the LBG correlation (solid blue) can be measured with signal-to-noise of roughly $1$ and $1.5\sigma$ in order of increasing redshift bin.  The width of each band in Fig.~\ref{fig:gy_error} indicates the redshift extent of that particular galaxy sample.

There is a clear evolution in the correlation function amplitude shown in Fig.~\ref{fig:gy_error} with redshift, corresponding to the changing statistical properties of the $y$ field over the course of reionization.   For reference, the top $x$-axis in Fig.~\ref{fig:gy_error} shows the evolution of the volume-averaged ionization fraction.  We caution that our fiducial forecasts have assumed a very high sensitivity CMB experiment, a factor of ten times better than the planned CMB-S4 sensitivity.  Furthermore, our assumption of $f_{\rm sky} = 0.5$ for the LAE sample may be unrealistic.  The light red and blue bands shown in Fig.~\ref{fig:gy_error} correspond to forecasts for CMB-S4 noise levels; in this case, the detection significance for the different cross-spectra is always below $2\sigma$.

\section{Discussion}
\label{sec:discussion}

We have shown that the correlation between high-redshift LAEs and LBGs with Compton-$y$ carries information about the reionization history of the Universe.  The galaxy-$y$ cross-spectra shown in Fig.~\ref{fig:spectra} reveal distinctive features at the scales of reionized bubbles, and show clear evidence for redshift evolution over the duration of reionization.  Similarly, Fig.~\ref{fig:gy_error} shows the redshift evolution of the average galaxy-$y$ power (over a restricted $\ell$ range), which roughly tracks the reionization history  of the Universe.  

Since the galaxy power is not expected to be shot-noise limited over the relevant angular scales ($\ell \lesssim 2000$, see Fig.~\ref{fig:spectra}), it is noise in the Compton-$y$ map that largely determines our ability to measure the LBG/LAE-$y$ cross-spectrum.  Both instrumental noise and $y$ from low-$z$ halos are important sources of noise for our analysis.  Masking halos with $M > 3\times 10^{13}\,M_{\odot}$ reduces the low-$z$ $y$ power by roughly two orders of magnitude, as seen in Fig.~\ref{fig:y_noise}.  In principle, less massive halos could be masked in order to further reduce $C_{\ell}^{yy-{\rm lowz}}$ and increase the signal-to-noise of the cross-correlation measurements.  However, as seen in Fig.~\ref{fig:y_noise}, masking halos much less massive than $3\times 10^{13}\,M_{\odot}$ will cause the instrumental and foreground noise to dominate the total noise power, and further halo masking will therefore be ineffective.  

Unfortunately, the amplitude of the $y$-LBG/LAE cross-spectrum is small compared to current and future noise levels in Compton-$y$ maps.  For expected CMB-S4 noise levels, no significant detection of the $y$-LBG/LAE cross-spectrum is expected.  A futuristic survey with ten times the sensitivity of CMB-S4, or roughly four times the sensitivity of PICO, could make a moderate-significance detection of the redshift evolution of the LAE-$y$ correlation, as shown in Fig.~\ref{fig:gy_error}, assuming a wide-field LAE sample.  By the time any such CMB mission is completed, however, we will likely have tighter constraints on reionization from 21 cm observations and from measurements of the kinematic Sunyaev-Zel'dovich effect.

Finally, we comment that our analysis has ignored potential systematics that might be important in very low-noise Compton-$y$ maps, such as contamination of $y$ maps by e.g. the cosmic infrared background (CIB).  Since the CIB is sourced primarily from $z  \lesssim 4$ \citep[e.g.][]{Schmidt:2014}, and because the morphology of any contaminating CIB signal is likely very different from the $y$ signal generated by the ionized bubbles, CIB contamination is unlikely to bias the $\lbgy$ measurements considered here, but it could act as a significant source of noise.  However, given our conclusion that even very futuristic surveys have little chance of detecting $\lbgy$ at high significance, we postpone a more careful consideration of such systematics to future work.

\section*{Acknowledgements}

We thank Adam Lidz for useful discussions and comments on our draft, and Colin Hill for help with the Compton-$y$ noise estimates.

LHW is supported by the Science and Technology Facilities Council (STFC). 

EJB and VI acknowledge support of the Kavli Foundation.

GK is partially supported by the Max Planck Society.

SP is supported in part by the U.S. National Science Foundation award AST-1440226.

Support by ERC Advanced Grant 320596 `\emph{The Emergence of Structure During the Epoch of Reionization}' is gratefully acknowledged.

We acknowledge PRACE for  awarding  us  access  to  the  Curie  supercomputer,  based in France  at  the Tr\'{e}s  Grand  Centre  de  Calcul  (TGCC). This work was performed using the Cambridge Service for Data Driven Discovery (CSD3), part of which is operated by the University of Cambridge Research Computing on behalf of the STFC DiRAC HPC Facility (\url{www.dirac.ac.uk}). The DiRAC component of CSD3 was funded by BEIS capital funding via STFC capital grants ST/P002307/1 and ST/R002452/1 and STFC operations grant ST/R00689X/1. DiRAC is part of the National e-Infrastructure.

The analysis code used in this work was written in \texttt{Rust} \citep[\url{https://www.rust-lang.org/},][]{Matsakis:2014:RL:2692956.2663188} and \texttt{Python} \citep[\url{https://www.python.org/},][]{van1995python}. In particular we employed the \texttt{SciPy}
\citep[\url{https://www.scipy.org/},][]{SCIPY} ecosystem of libraries including: \texttt{NumPy}
\citep[\url{https://www.numpy.org/},][]{NUMPY}, \texttt{Matplotlib}
\citep[\url{https://matplotlib.org/},][]{MATPLOTLIB} and \texttt{Cython}
\citep[\url{https://cython.org/},][]{CYTHON}. 

\section*{Data availability}

The data used to generate the figures in this work are available upon request.

\nocite{Toshiya:inprep}

\bibliographystyle{mnras}
\bibliography{thebib}

\appendix
\apptablenumbers
\section{The hot halo gas contribution and the effect of feedback}
\label{sec:feedback}
In this work we are chiefly concerned with the correlations between high-redshift galaxy populations and the $y$-signal arising from the IGM during reionization. As discussed in Section~\ref{sec:noise}, a key component in the low-redshift $y$-signal comes not from the IGM but from the hot gas within halos (i.e. from clusters) where stellar feedback is important. In this appendix we explore to what extent hot halo gas contributes to the high-redshift $y$-signal, and also whether the simulation of feedback impacts our results.

\begin{figure*}
    \centering
    \includegraphics[width=0.48\textwidth]{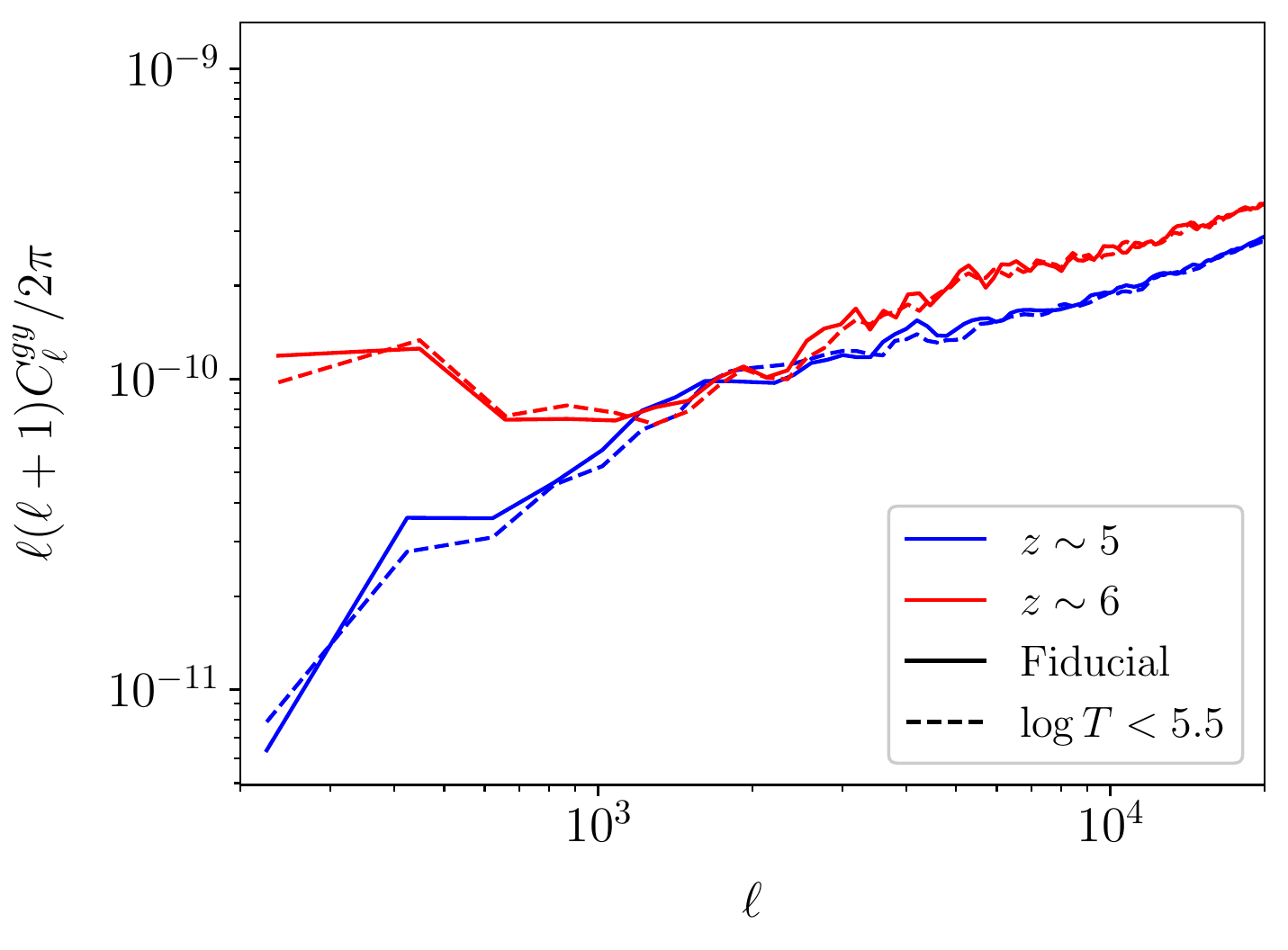}
    \includegraphics[width=0.48\textwidth]{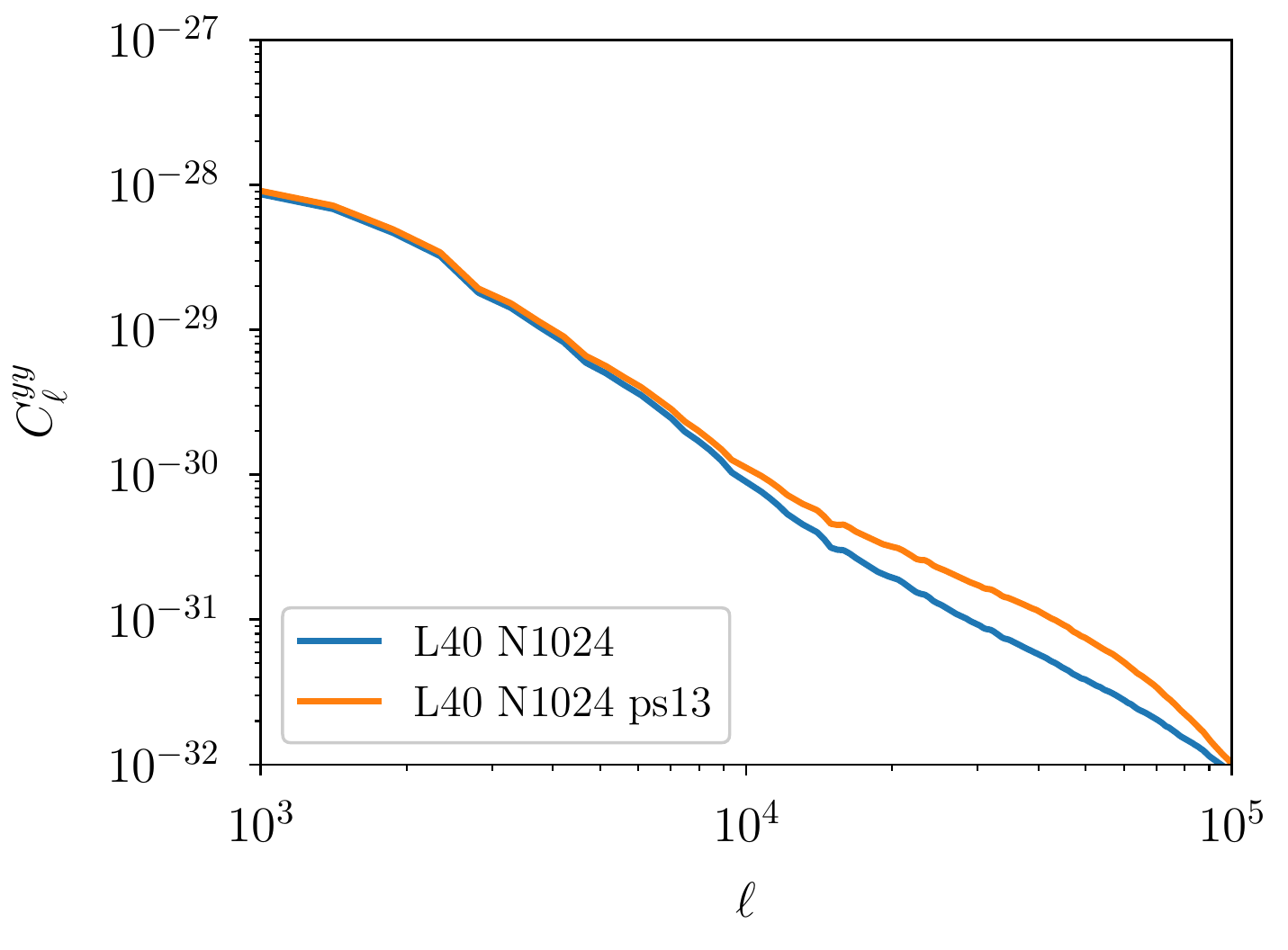}
    \caption{
    Testing the contribution of hot halo gas and the effects of feedback.
    \emph{Left}: A comparison of the dimensionless y-LBG cross-power spectrum for our fiducial simulations (solid lines) and for the case when the temperature in the simulation is capped at $T < 10^{5.5}$ K (dashed lines).
    \emph{Right}: The $y$ auto-spectrum at $z=7$ for two simulations: ``{\tt L40 N1024}'' (blue) which uses the fiducial feedback prescription employed elsewhere in this work, and ``{\tt L40 N1024 ps13}'' (orange) which uses the supernova-driven feedback model of \citet{PUCHWEIN13}.
    }
    \label{fig:feedback}
\end{figure*}

To test the contribution of hot halo gas, we recalculate the $y$-signal with an upper temperature limit of $10^{5.5}$ K, setting all gas above this temperature to this cap. In the left panel of Figure~\ref{fig:feedback} we show a comparison using our fiducial $L=160$ cMpc/h simulation of the effect of introducing this temperature cap. In this panel we compare the y-LBG cross-power spectra, and find that introducing a temperature cap does not significantly affect our results. This suggests that the hot halo gas is not a dominant contribution to the large-scale correlation signal, but that as discussed in Appendix~\ref{sec:calculation}, we expect the signal to arise predominantly from the ionised bubble distribution.

Regarding feedback, the Sherwood simulations we employ in this work were designed to capture the evolution of the low-density IGM, and approximate star formation within halos using the {\tt QUICK\_LYA} feedback prescription \citep[see Section~\ref{sec:simulations},][]{VIEL04}. 
To test the effect of feedback we employ two further simulations from the Sherwood suite with box sizes $L=40$ cMpc/h and particle numbers $N=2\times1024^3$: (i) a fiducial simulation run using {\tt QUICK\_LYA} feedback prescription (referred to as~{\tt L40 N1024}); (ii) a second simulation run with a more accurate feedback prescription based on the supernova-driven wind model of \citet{PUCHWEIN13} (referred to as~{\tt L40 N1024 ps13}). In the right panel of Figure~\ref{fig:feedback} we show the angular $y$ auto-power spectra for both $L=40$ cMpc/h simulations to compare the impact of feedback. We see that whilst there is some deviation on small scales $\ell > 10^4$, there is good agreement in the large scale regime which is of central interest to this work. This suggests that our results are robust to the choice of feedback prescription.

\section{Analytical calculation}
\label{sec:calculation}

In this appendix we provide rough analytical estimates for the expected $C^{gy}_{\ell}$ signal discussed in the main text.  The $y$ signal is given by 
\begin{equation}
y = \frac{\sigma_T}{m_e c^2} \int \dd l \, P_e(l),
\end{equation}
where $l$ is physical distance along the line of sight and $P_e$ is the electron pressure.  We consider galaxies located in a redshift bin that is is not much wider than the size of the ionized bubbles.  This assumption simplifies the calculation because it allows us to ignore the impact of multiple ionized bubbles along the line of sight.  For the case of the LAEs --- which are selected with narrowband observations --- this assumption is quite reasonable; for the LBGs, this assumption is likely to be less accurate.  We make the approximation that all galaxies live at the centers of reionized bubbles.  Finally, we assume that the ionization fraction and pressure within the bubble are constant (not an unreasonable approximation, see e.g. \citealt{WEINBERGER18}).  Making these approximations, the $y$ along the direction to a galaxy becomes
\begin{eqnarray}
y_{\rm bubble} = \frac{2\sigma_T R P_e}{m_e c^2(1+z)},
\end{eqnarray}
where $R$ is the bubble radius in comoving units.  We can write the pressure in the bubble as
\begin{eqnarray}
P_e \sim k_B (1+\delta)\Omega_b \rho_c(1+z)^3 \left(X_p(1/u) + Y_p(2/4u) \right)T,
\end{eqnarray}
where $\delta \sim 4$ is the overdensity in a bubble, $T$ is the bubble temperature and $u$ is the atomic mass unit.  Assuming 
$T \sim 10^4\,{\rm K}$ we have
\begin{eqnarray}
y_{\rm bubble} \sim 1.7\times 10^{-11} (1+\delta) (1+z)^2 \left(\frac{R}{10\,{\rm cMpc/h}}\right).
\end{eqnarray}
At $z = 7$, and for $\delta = 4$, we find $y \sim 5.6\times 10^{-9}$ for a bubble of radius $10\,{\rm cMpc/h}$.  This value agrees well with the LAE results seen in the Sherwood simulations (Fig.~\ref{fig:y_maps}).  

We now consider estimates for $C^{gy}_{\ell}$.  The multipole corresponding to the bubble size is roughly
\begin{eqnarray}
\ell_{\rm bubble} \sim \pi/\theta_{\rm bubble} \sim \pi \chi(z)/R.
\end{eqnarray}
From Fig.~\ref{fig:y_maps}, at $z \sim 7$ we have $R \sim 10\,{\rm cMpc/h}$, so $\ell_{\rm bubble} \sim 1900$; at $z \sim 6.6$ we have $R \sim 30\,{\rm cMpc/h}$, so $\ell_{\rm bubble} \sim 640$.   These values agree well with the location of the upturns of $C^{gy}(\ell)$ seen in Fig.~\ref{fig:spectra}.  This suggests that the shape of the cross-spectra directly encode information about the bubble sizes.

The cross-power is related to the $y$-profile of the bubble via a Hankel transform:
\begin{eqnarray}
C^{gy}_{\ell} = 2 \pi \int \dd \theta \, \theta J_0(\ell \theta)  \xi^{gy}(\theta),
\end{eqnarray}
where $\xi^{gy}(\theta)$ is the bubble profile as a function of angular separation, $\theta$.  Still assuming that all galaxies live at the centers of equally sized bubbles, and that the $y$ profile inside a bubble is constant, we have
\begin{eqnarray}
C^{gy}_{\ell} \sim 2 \pi \int_0^{\theta_{\rm bubble}} \dd \theta \, \theta J_0(\ell \theta) y_{\rm bubble},
\end{eqnarray}
where $\theta_{\rm bubble}$ is the angular scale corresponding to the radius of the bubble at its redshift.  Performing this integral for the bubble radius and $y$ amplitude estimated above at $z \sim 7$, the predicted amplitude of the cross-power is $C^{gy}(\ell = 2000) \sim 8\times 10^{-15}$.

Some bubbles will of course be much smaller than $10\,{\rm cMpc/h}$.  For these bubbles, the $y$ amplitude at $R$ will be zero, and the $\xi^{gy}$ correlation at $R$ will therefore be suppressed.  If we assume that a fraction $\beta$ of galaxies live in bubbles smaller than $10\,{\rm cMpc/h}$, then the correlation function at this scale will be suppressed by $(1-\beta)$.   The true measured cross-power is roughly $C^{gy}(\ell = 2000) \sim 2\times 10^{-15}$.  This suggests that $(1-\beta) \sim 0.25$.  From  Fig.~\ref{fig:y_maps}, at $z \sim 7$ it seems reasonable that roughly 75\% of galaxies live in bubbles smaller than $10\,{\rm cMpc/h}$.  Our analytic estimates of the cross-power and the simulation results therefore appear roughly consistent.

\bsp	
\label{lastpage}
\end{document}